# Parsimonious Network based on Fuzzy Inference System (PANFIS) for Time Series Feature Prediction of Low Speed Slew Bearing Prognosis

Wahyu Caesarendra[*a,d], Mahardhika Pratama[b], Tegoeh Tjahjowidodo[c], Kiet Tieu[d], and Buyung Kosasih[d]

[a] Mechanical Engineering Department, Diponegoro University, Tembalang, Semarang 50275, Indonesia
[b] School of Computer Science and Engineering, Nanyang Technological University, Singapore 639798
[c] School of Mechanical and Aerospace Engineering, Nanyang Technological University, Singapore 639798
[d] School of Mechanical, Materials, Mechatronic and Biomedical Engineering, University of Wollongong, Wollongong, New South Wales 2522, Australia

*Corresponding Author:

Dr. Wahyu Caesarendra

Department of Mechanical Engineering

Diponegoro University Indonesia

E-mail: wc026@uowmail.edu.au

Total number of pages: 35

Total number of figures: 8

Total number of tables: 4

# Parsimonious Network based on Fuzzy Inference System (PANFIS) for Time Series Feature Prediction of Low Speed Slew Bearing Prognosis

**Abstract**

In recent years, the utilization of rotating parts, e.g. bearings and gears, has been continuously supporting the manufacturing line to produce consistent output quality. Due to their critical role, the breakdown of these components might significantly impact the production rate. A proper condition based monitoring (CBM) is among a few ways to maintain and monitor the rotating systems. Prognosis, as one of the major tasks in CBM that predicts and estimates the remaining useful life of the machine, has attracted significant interest in decades. This paper presents a literature review on prognosis approaches from published papers in the last decade. The prognostic approaches are described comprehensively to provide a better idea on how to select an appropriate prognosis method for specific needs. An advanced predictive analytics, namely Parsimonious Network Based on Fuzzy Inference System (PANFIS), was proposed and tested into the low speed slew bearing data. PANFIS differs itself from conventional prognostic approaches in which it supports for online lifelong prognostics without the requirement of retraining or reconfiguration phase. The method is applied to normal-to-failure bearing vibration data collected for 139 days and to predict the time-domain features of vibration slew bearing signals. The performance of the proposed method is compared to some established methods such as ANFIS, eTS, and Simp_eTS. From the results, it is suggested that PANFIS offers outstanding performance compared to those of other methods.

## 1. Introduction

Prognosis methods are commonly applied to predict the lifetime of rotating components, which generally can be divided into two stages. The first stage refers to the normal zone where no significant deviation from the normal operating state observed. The second stage is abnormal zone; this stage is initiated by potential failure that progressively develops into actual failure [1]. It is on the second stage the prognosis methods are usually applied to predict unexpected failures in time basis from the incipient or impeding damage using either event data or condition monitoring (CM) data.

Since more and more industries today move towards the so-called industry 4.0, more reliable predictive analytics is required. This requirement forces industries to install more sensors for capturing some process variables which results in an exponential increase of problem space. In such situation, conventional predictive analytics are hardly scalable [2], while a quick decision based on the most recent observation is typically needed. This requirement is becoming difficult to fulfil without an online real-time predictive analytics that is capable of delivering a reliable prediction while updating the model to keep pace with high-speed processes in typical manufacturing plants. Most processes are also subjected to a number of rapidly changing external variables. This trait cannot be handled by a static model where its structure is fully determined in its initial design. A model is supposed to be flexible for new concepts which normally lead to expansion of its initial structure. An over-complex structure adversely affects model's generalization because of overfitting. These research issues have led to algorithmic development of the so-called evolving intelligent systems (EIS) [3, 4], which have attracted significant research interest since the past decade [5-11]. EISs have been successfully deployed in several predictive maintenance tasks [12-14].

This paper aims to present two research studies: (1) a literature review on prognosis approaches from published papers in the last decade. The prognostic approaches are described comprehensively to provide a better idea on how to select an appropriate prognosis method for specific needs; (2) a time-series feature prediction of a low speed slew bearing prognosis using a seminal work, namely the parsimonious network based on fuzzy inference system (PANFIS) [11]. PANFIS features a fully open structure where its network structure can be self-evolved on the fly from data streams. PANFIS utilises the theory of statistical contribution [9, 10] to grow and prune its fuzzy rules. A unique feature of PANFIS is seen its generalized Takagi Sugeno Kang (TSK) fuzzy system where it scatters multivariate Gaussian functions as rule premise, while making use of the first-order TSK rule consequent. It is well-known that such rules obscure rule semantics since the atomic rules of classic fuzzy rules vanish. PANFIS is equipped by a transformation strategy which extracts fuzzy set representation of high-dimensional ellipsoidal cluster. The fuzzy set merging strategy is incorporated because projection of ellipsoids to one-dimensional space normally results in overlapping fuzzy sets. The parameter learning strategy is based on the Extended Recursive Least Square method [8], which appends a binary function to enhance convergence and stability of tuning process.

In order to select an appropriate prognosis method for slew bearing, a review of existing prognosis methods is first presented in this chapter. The review consists of two parts: (1) a brief review on classification of prognosis approaches and (2) a review on prognosis methods for rolling element bearings.

## 2. Classification of prognosis approaches

There exists in literature some reviews on the prognosis approaches for rotating machineries [15-22]. For example, Lee et al. [15] reviewed prognosis methods for critical components such as bearing, gear, shaft, pump and alternator. The authors classified the prognosis approaches into three, namely model-based, data-driven and hybrid prognosis approaches. However, the classification does not include complete methods on each prognosis approach. In their work, two methods are classified as a model-based approach, i.e. alpha-beta-gamma tracking filter and Kalman filter, while neural network (NN), fuzzy logic and decision tree are classified as a data-driven approach.

Another review paper by Jardine et al. [16] presented a clearer classification in prognosis approaches, but the review emphasized more on rotating machinery diagnostics rather than machinery prognostics. The authors classified the prognosis approaches into three groups: statistical, artificial intelligent (AI) and model-based approaches. The statistical approaches include statistical process control (SPC), logistic regression, autoregressive and moving average (ARMA), proportional hazard model (PHM), proportional intensity model (PIM) and hidden Markov model (HMM). In AI techniques, e.g. artificial neural network (ANN) and its sub-classes such as self-organising neural networks, dynamic wavelet neural networks and recurrent neural networks, back propagation neural network and neural-fuzzy inference systems are still commonly used in AI prognostics. Among model-based approaches, defect propagation models via mechanistic modelling and crack growth rate model are the commonly used methods.

A popular review paper on prognosis methods was presented by Heng et al. [20]. The authors classified the methodologies for predicting rotating machinery failure into two different groups, namely physics-based and data-driven prognosis models. A number of papers focusing on physics-based prognostics which used Paris' formula are still found to be dominant [20]. Other methods such as finite element analysis (FEA) to calculate stress and strain field, and Forman law of linear elastic fracture mechanics are also classified in physics-based approaches. Similar to the result from the two review papers previously mentioned, ANN and its variants is currently the most commonly used methods in the data-driven prognosis class. Other methods such as fuzzy logic, regression analysis, particle filtering, recursive Bayesian technique and HMM are also included in data-driven prognosis methods.

However, a number of methods within the data-driven methods as presented in [20] need further sub-classification in terms of artificial intelligent or statistical approaches.

A recent review is presented by Lei et.al. [23], which mentioned that a machinery prognostic method generally consists of four technical processes, i.e. data acquisition, health indicator (HI) construction, health state (HS) division, and RUL prediction. In addition, they also explained that the existing research work and literature review have converged to the four processes, in particular especially the latter one. The paper presents a systematic review that covers the four technical processes comprehensively.

A list of literatures which reviewed the prognosis approaches is presented in Table 1. The table provides the year of publication and the classification groups of prognosis approaches. Although the merits and the demerits of prognosis approaches have also been presented in Lee et al. [15] and Heng et al. [20], to date there have been scientific gaps of prognostics reviews, namely (1) *the classification of prognostics approaches that remain unclear*: many review papers presented different classifications as seen in Table 1 and sometimes the classified methods in each approach are overlapping depending on the authors; and (2) *in what applications can certain prognosis approach be used*: the prognosis literature provides few information to help typical industry users in selecting an appropriate approach or method for their specific needs. This paper aims to bridge the gap by providing another classification of prognosis approaches. Such approaches include the basic prognosis method selection for specific needs (shown in Figure 1) and the classified methods of each prognosis approach as presented in Section 2.

**Table 1** Classification of prognosis approaches extracted from literatures.

| No | Author | Refs. | Year | Classification groups |
|----|--------|-------|------|----------------------|
| 1 | Lei et al. | [23] | 2018 | - Physics model-based approaches<br>- Statistical model-based approaches<br>- AI approaches<br>- Hybrid approaches |
| 2 | Kan et al. | [24] | 2015 | Data-driven approaches:<br>- Statistical techniques<br>- AI techniques |
| 3 | Lee et al. | [15] | 2014 | - Model-based<br>- Data-driven<br>  Hybrid prognostics approaches |

**Table 1** Classification of prognosis approaches extracted from literatures. (Cont'd)

| No | Author | Refs. | Year | Classification groups |
|---|---|---|---|---|
| 4 | Jardine et al. | [16] | 2006 | - Statistical approaches<br>- Artificial intelligent approaches<br>- Model-based approaches |
| 5 | Kothamasu et al. | [17] | 2006 | - Reliability-based<br>- Model-based<br>- Signal-based<br>- Statistical |
| 6 | Goh et al. | [18] | 2005 | - Data-driven<br>- Model-based |
| 7 | Vachtsevanos et al. | [19] | 2006 | - Data-driven<br>- Experience-based<br>- Model-based (Physics of failure) |
| 8 | Heng et al. | [20] | 2009 | - Physic-based prognostics models<br>- Data-driven prognostics models |
| 9 | Peng et al. | [21] | 2010 | - Physical model-based methodology<br>- Knowledge-based methodology<br>- Data-driven methodology<br>- Combination model |
| 10 | Dragomir et al. | [22] | 2009 | - Model-based approaches<br>- Data-driven approaches |
| 11 | Hines et al. | [25] | 2008 | - Time-to-failure data-based prognostics<br>- Stress-based prognostics<br>- Effect-based prognostics |
| 12 | Kim | [26] | 2010 | - Data-driven approaches<br>- Model-based approaches<br>- Reliability-based approaches |

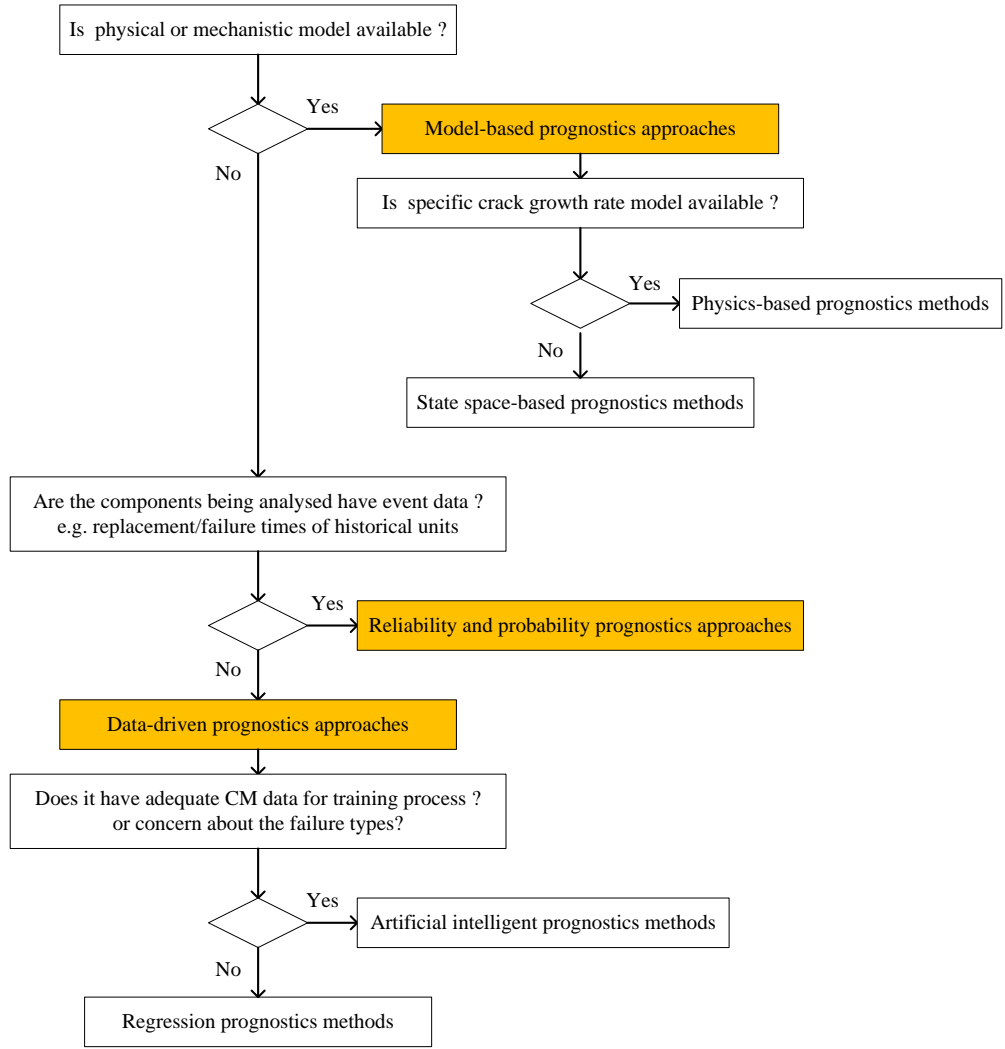

**Figure 1.** A general prognosis approach or method selection for specific needs.

2.1 Prognosis methods for rolling element bearings

According to the literature review presented in Table 1, four prognosis approaches have been adopted in this paper. The prognosis methods of each approach are reviewed and presented in Table 2. Table 2 also presented the degradation parameters or features extractions which were applied in a particular method or algorithm. It can be seen that RMS and kurtosis are the most commonly used features in prognosis methods. It is worth noting that the reviewed methods in this paper only focused on rolling element bearings prognostics. The approaches are: (1) model-based approaches; (2) reliability-based methods and probability models; (3) data-driven approaches; and (4) combined data-driven approaches and reliability-based methods.

**Table 2** Prognosis methods for rolling element bearing.

| No | Classification | Method or algorithm | Features used |
|---|---|---|---|
| 1 | Model-based approaches<br>- Physics-based<br>  prognostics models<br><br>- State space-based<br>  Methods | - Paris' formula [27, 28]<br>- Stiffness-based prognostics<br>  model [29]<br>- Kalman filter [30]<br>- Particle filter [31] | N/A<br>N/A<br><br>N/A<br>RMS and envelope<br>acceleration |
| 2 | Reliability-based<br>methods and<br>probability models | - Gaussian process models [32]<br>- PIM [33]<br>- PCM [34]<br>- Stochastic model [35]<br>- PHM [36, 37]<br>- Weibull distribution [[38]]<br>- HMM [39]<br>- WPD and HMM [40] | Rényi entropy<br>Kurtosis<br>Principal features<br>N/A<br>N/A<br>N/A<br>RMS<br>Peak-to-peak, energy<br>and kurtosis |
| 3 | Data-driven approaches<br>- Artificial intelligence (AI)<br>  methods<br><br>- Regression methods<br><br><br><br><br>- Combined AI and<br>  regression methods | - ANN [41, 42]<br>- Fuzzy logic [43]<br>- Genetic algorithm (GA) [44]<br>- ALE and ARIMA [45]<br>- Recursive least square (RLS) [28]<br>- Dempster-Shafer regression [46]<br><br>- ARMA/GARCH model [47]<br><br>- RVM and logistic regression [48]<br>- RVM and exponential regression [49] | N/A<br>N/A<br>Monitoring index<br>RMS, skewness, kurtosis<br>RMS<br>RMS and envelope<br>acceleration<br>RMS and envelope<br>acceleration<br>Kurtosis<br>RMS |
| 4 | Combine data-driven and<br>reliability-based methods | - Cox-proportional hazard (CPH)<br>  model and SVM [36]<br>- ARMA, PHM and SVM [37]<br><br>- SVM and survival probability [50]<br>- RVM and survival probability [51]<br>- ANN and Weibull distribution [52] | Kurtosis<br><br>Peak acceleration<br>and RMS<br>Kurtosis<br>Kurtosis<br>RMS, kurtosis and<br>entropy estimation |

2.2 Review on model-based approaches

Methods included in model-based approaches require an accurate mathematical model to be developed. They also use residuals as features, where residuals are the outcomes of consistency checks between the measurements of a real system and the outputs of a mathematical model [26]. Model-based prognosis approaches in this paper are divided into two classes, namely physics-based prognosis models and state space-based methods.

*2.2.1 Physics-based prognostics models*

The physics-based approaches assume that accurate mathematical or physical models of the monitored system or machine are available and provide a technically comprehensive approach that has been used traditionally to understand failure mode progression. A common model used in physics-based prognosis model is Paris's law equation. The physics-based methods are able to predict the failure progression accurately if the appropriate model is used. However, a limitation of physics-based approaches is associated to its inflexibility which means that the particular model can only be applied to specific types of components [53].

Some literatures applied model-based prognosis methods are presented as follows:

Li et al. [27] present a defect propagation model by mechanistic modelling approach for bearing prognosis to estimate RUL of rolling element bearing.

In slew bearing prognosis study, several works has been conducted. Potočnik et al. [54] calculate the maximal contact force by means of analytical expression of the Hertzian contact theory, and then used a strain-life model to calculate the fatigue life on the basis of the subsurface stresses. Glodez et al. [55] compare the two methods for calculating the fatigue life of a slewing bearing: strain-life approach and stress-life approach based on ISO 281 [56]. Results show that the stress-life approach is the most precise method for calculating bearing fatigue lifetime.

*2.2.2 State space-based methods*

Besides physics-based methods, the state space-based methods such as Kalman filter and particle filter are also considered as a part of model-based prognosis methods because it builds a dynamic model of system being analysed to predict a future point in time.

Kalman filtering (KF) incorporates the signal embedded with noise and forms that can be considered as a sequential minimum mean square error estimator (MMSE) of the signal [26].

Particle filter or Monte Carlo methods for nonlinear filtering are based on sequential versions of the importance sampling paradigm. This is a technique that amounts to simulating samples under an instrumental distribution and then approximates the target distributions by weighting these samples using appropriately defined importance weight. Particle filter offers the great advantage of not being subject to the assumption of linearity, Gaussianity and stationary [31]. Particle filter for rolling element bearing prognostics is presented in [31].

2.3 Review on reliability-based methods and probability models for prognosis

As aforementioned, prognostics is used to predict how much time left before a failure occurs given the current machine condition and past operation profile which is commonly called as remaining useful life (RUL). In some situation, especially when a fault is catastrophic (e.g., nuclear power plant), it would be desirable to predict the chance that a machine operates without a fault or a failure up to some future time given the current machine condition and past operation profile [16]. This issue can be addressed using failure-based reliability or probability prognosis models. Failure-based reliability is used to estimate the lifetime distribution and its parameters when sufficient, complete and/or censored failure time data exist. If prior knowledge of the lifetime distribution exists for similar components, then often the lifetime distribution is assumed to follow the same distribution of a similar component [26]. For example, Goode et al. [57] separate two intervals of whole machine life: the I-P (Installation-Potential failure) interval in which the machine is running normally and the P-F (Potential failure-Functional failure) in which the machine condition has a problem. Based on two Weibull distributions assumed for the I-P and P-F intervals, failure prediction has been derived in the two intervals and the RUL is estimated.

*2.3.1 Proportional hazard model*

Proportional hazards models (PHMs) are commonly used in failure prediction and reliability analysis. The method was proposed by Cox in 1972 [58] and was first introduced in the clinical studies to characterise the disease progression in existing cases by revealing the importance of covariates [36]. It is the most popular model for survival analysis due to its simplicity. The reason is that it is not based on any assumptions concerning the nature or shape of the survival distribution [36]. PHMs assume that hazard changes proportionately with covariates and that the proportionality constant remains the same at all time [20]. The method has been used in bearing prognostics [36, 37]. A review of the existing literature on the PHM is presented in [59]. Usually PHM cannot be used as a stand-alone prognostic method. PHM is usually used together with AI method. PHM is used to build the degradation model, based on this model, AI method e.g. SVM is used to predict the degradation model [36, 37].

*2.3.2 Proportional covariates model*

A proportional covariates model (PCM) is proposed by Sun et al. [34]. PCM can be used to estimate the hazard functions of mechanical components in cases of sparse or no historical failure data provided that the covariates are proportional to the hazard.

*2.3.3 Reliability model*

Heng et al. [60] introduce an intelligent reliability model called the intelligent product limit estimator (PLE), which was able to include suspended CM data in machinery fault prognosis. The accurate data modelling of suspended data has been found to be of great importance, since in practice machines are rarely allowed to run to failure and data are commonly suspended. The model consists of a feed-forward neural network (FFNN) whose training targets are asset survival probabilities estimated using a variation of the Kaplan-Meier estimator and the true survival status of historical units.

*2.3.4 Proportional intensity model*

Vlok et al. [33] utilise statistical residual life estimate (RLE) on roller bearings to study changes in diagnostics measurements of vibration and lubrication levels which can influence bearing life. RLEs are based on proportional intensity models (PIMs) and mainly used for non-repairable systems utilising historic failure data and the corresponding diagnostic measurements.

*2.3.5 Stochastic model*

A prediction method for residual life of rolling element bearing based on stochastic process called gamma process is presented in [35].

*2.3.6 Weibull distribution*

In slew bearing cases, several reliability methods for prediction have also been studied. Yang et al. [38] present the reliability prediction approach for slew bearing based on the Weibull distribution. Hai et al. [61] develop a method for evaluating rolling contact fatigue (RCF) reliability of slew bearings, which replaced the reliability factor $a_1$ from ISO 281 with the Lundberg-Palmgren theory.

*2.3.7 Hidden Markov model*

The use of hidden Markov models (HMMs) in bearing fault prognosis is investigated by Zhang et al. [39]. In a HMM, a system is modelled to be a stochastic process in which the subsequent states have no causal connection with previous states [20]. It is assumed that the state transition time of estimated vectors follows some multivariate distribution. Once the distribution is addressed, the conditional probability distribution of a distinct state transition can be estimated [21].

Ocak et al. [40] develop a new robust scheme based on wavelet decomposition and hidden Markov model for tracking the severity of bearing faults, and reached the conclusion that the probabilities of the normal bearing's hidden Markov model keep decreasing as the bearing damage progresses toward bearing failure.

Tallian [62] presents a rolling bearing life prediction model using statistical lifetime determination.

2.4 Review on data-driven prognosis approaches

Data-driven approaches are derived directly from routine condition monitoring (CM) data of the monitored system (e.g. temperature, vibration, oil debris, current, etc). Data-driven approaches can be regarded as degradation-based methods because they focus on using measures of component degradation, not on failure data, to assess the remaining of a component. In other words data-driven approaches rely on the availability of run-to-failure data and require performing suitable extrapolation to the damage progression to estimate RUL. One major advantage of these techniques is the simplicity of their calculations [20] because these methods do not require mechanistic or physical knowledge of the system or component being analysed. Data-driven approaches may often produce more available solution in many practical cases. The reason is probably that the data-driven models calculated from data-driven methods are easier to obtain compared to an accurate model from a system or component. A main drawback of data-driven approaches is their dependency on the equality of the monitored data. In data-driven approaches, the proper selection of a trending parameter or feature is the key issue in implementing the prognosis. The selection criteria for such parameter should include the diagnosis ability, sensitivity, consistency and the amount of calculation required [41]. In this paper, data-driven approaches are divided into three sub-categories: (i) artificial intelligence (AI) methods; (ii) regression methods; and (iii) combined AI and regression methods.

*2.4.1 AI methods*

The first data-driven approaches sub-class (AI methods) are based on machine-learning techniques for prognostics. AI methods predict the selected features that correlate with the failure progression based on the learning or training process. The methods rely on past patterns of degradation to project future degradation. The features used in AI methods are extracted from CM data e.g. vibration signals. More CM data are used in the training process, and more accurate model is obtained, but computational time increases. As AI methods use experimental data to train the methods in order to build a prediction model, thus, AI methods are highly-dependent on the quantity and quality of the measured data. In general, AI methods adopt a one-step or multi-step ahead prediction technique in order to predict the future state. A review of AI methods for prognostics can be found in [63]. Several AI methods have been developed for decades. Artificial neural network (ANN) and its variants such as self-organizing map (SOM) and back propagation neural network (BPNN) methods are most commonly used [41, 42, 64-66]. Although ANN is the most commonly used method and it has worked successfully in bearing prognosis application, it has fundamental drawbacks in model development. Such drawbacks include how many hidden layers should be included and what is the number of processing nodes that should be used for each layer. These are the major questions for users.

Another popular AI technique that is used for prognostics is the fuzzy logic technique. Fuzzy logic provides a language (with syntax and local semantics) into which one can translate qualitative knowledge about the problem to be solved. In particular, fuzzy logic allows the use of linguistic variables to model dynamic systems. These variables take fuzzy values that are characterized by a sentence and a membership function. The meaning of a linguistic variable may be interpreted as an elastic constraint on its value. These constraints are propagated by fuzzy inference operations. The resulting reasoning mechanism has powerful interpolation properties that in turn give fuzzy logic a remarkable robustness with respect to variations in the system's parameters and disturbances.

Other AI prognostics methods for bearing have been applied e.g. LVQ is used by Zhang et al. [67] to generate a sequence of codes for representing fault signatures in the model.

*2.4.2 Regression methods*

The second data-driven sub-class (regression methods) are based on time series analysis techniques for prognostics. The regression methods are useful if a reliable or accurate system model is not available. The data-driven prognosis approaches through regression methods are used to determine the RUL. This is achieved by trending the trajectory of a developing fault and predicting the amount of time before it reaches a predetermined threshold level [15].

Niu and Yang [68] introduce the Dempster-Shafer regression for multi-step-ahead prediction of a methane compressor in a petrochemical plant. Using the similar vibration data, Pham et al. [47] develop a forecasting method based on ARMA/GARCH model.

Kosasih et al. [45] present the adaptive line enhancer (ALE) and auto-regressive integrated moving average (ARIMA). Jantunen [43] uses high-order regression functions to mimic bearing fault development and also to save trending data in a compact form.

*2.4.3 Combined AI and regression methods*

In combination of AI and regression methods, Caesarendra et al. [48] develop a combined prognostics method based on regression and AI method. Logistic regression (LR) is used to estimate failure degradation of bearing based on run-to-failure datasets and the results are then regarded as target vectors of failure probability. Relevance vector machine (RVM) is then used to train the run-to-failure bearing data and target vectors of failure probability. After the training process, RVM is employed to predict failure probability of individual bearing. The performance of the proposed method is validated using experimental and simulated data. The result shows the plausibility and effectiveness of the method which can be considered as the machine degradation assessment model. In another work, Caesarendra et al. [69] present a combined Cox-proportional hazard model and SVM. The failure rate is calculated using the Cox-PHM. The Kurtosis is extracted as a bearing condition parameter under specified operating conditions. SVM is trained using the kurtosis and the failure rate as a target vector to build the prediction model. The trained SVM is then used to predict the final failure time of individual bearing.

Tran et al. [70] employ a multi-step-ahead regression technique and ANFIS to predict the trending data.

2.5  Review on combined data-driven method and reliability-based methods

In this method, the statistical methods are used when the AI prognostics methods require quantitative data measurements.

Widodo and Yang [50] develop an intelligent machine prognostics method using survival probability method namely survival analysis (SA) and SVM. The method has the benefit that it employed censored data which is commonsensical in practice. SA utilizes censored and uncensored data collected from CM routine and then estimates the survival probability of failure time of machine components. SVM is trained by data input from CM histories data that correspond to target vectors of estimated survival probability. After validation process, SVM is employed to predict failure time of individual unit of machine component. Still dealing with censored data and survival probability analysis, Widodo and Yang [51] develop a combined survival probability analysis and RVM. In this study, Kaplan-Meier (KM) estimator is used to build a survival probability model. The RVM is then used to train the model and predict the final failure of individual unit of machine component.

## 3. Experimental setup

3.1 Slew bearing test-rig and data acquisition

The run-to-failure data used in this paper was collected from slew bearing test rig. The test rig was designed to replicate an actual condition in steel mill manufacturing that operate the bearing in low rotational speed, high load and dust environment. Figure 2 shows the schematic of the slew bearing test rig including the main drive gear reducer, the hydraulic load and how the bearing is attached. A detailed sensors placement is presented in Figure 3. Four accelerometers, two AE sensors and four temperature sensors were used during the experiment. Two accelerometers of IMI 608A11 ICP type sensors with sensitivity 100mV/g and frequency range 0.5 to 10 kHz, and two accelerometers of IMI 626B02 ICP type sensors with sensitivity 500mV/g and frequency range 0.2 to 6 kHz were used. The IMI 608A11 ICP type sensors were installed on the inner radial surface at 180 degrees to each other and the IMI 626B02 ICP type sensors were attached on the axial surface at 180 degrees to each other. Similar to the measurement in continuous rotation, these accelerometers were connected to a high speed Pico scope DAQ (PS3424). The IMI 626B02 ICP type accelerometers were selected because of the minimum frequency range of 0.2Hz and because the sensitivity is higher than that of the IMI 626B02 ICP type accelerometer. The vibration signal was acquired using 4880 Hz sampling rate.

A three axes (two axial rows and one radial row) brand new slew bearing is used in this experiment. Each axial and radial row has dozens of rollers inside. The slew bearing is typically large in dimension and is usually used to support high axial and radial load [71]. The bearing attachment to the test rig is shown in Figure 2(b); and 30 tonne was applied to the bearing. The vibration signal of the bearing was acquired two times daily (morning and night). To accelerate the failure, coal dust contamination was injected to the bearing on day 90.

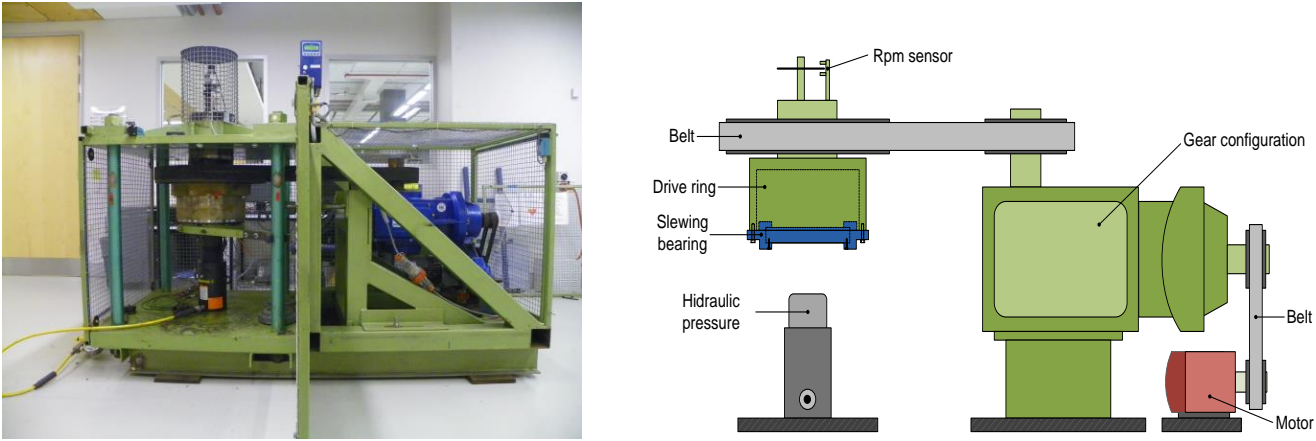

**Figure 2.** (a) Slew bearing rig picture (b) Schematic of laboratory slew bearing rig showing a slew bearing attached in the drive ring and the applied load from hydraulic.

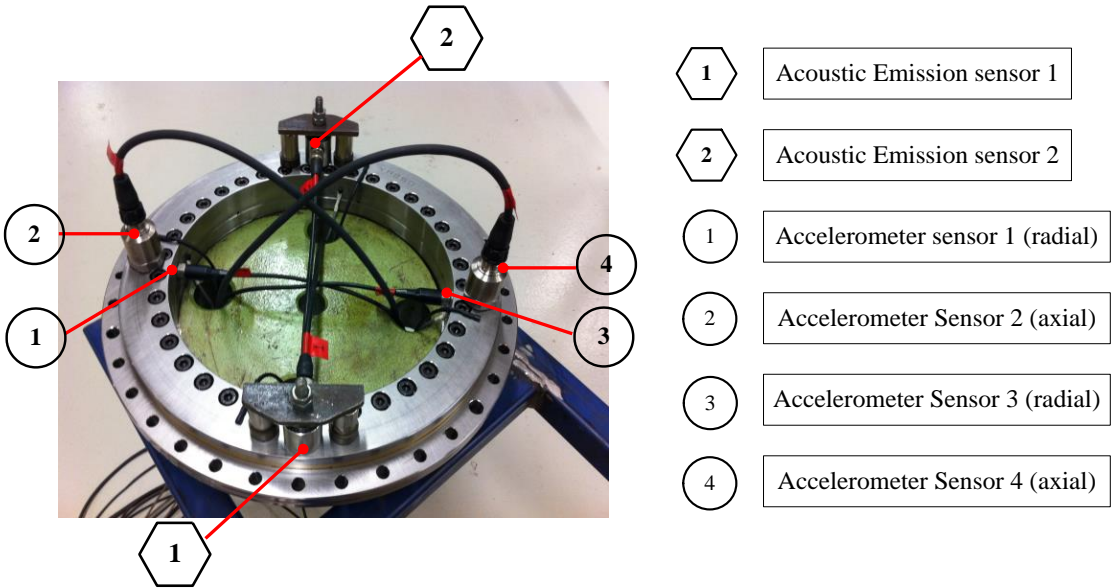

**Figure 3.** A detailed sketch of accelerometers and AE sensors location.

3.2 Feature extraction

Nine time-domain features (i.e. RMS, variance, skewness, kurtosis, shape factor, crest factor, entropy, histogram upper and histogram lower) are extracted from four vibration data collected in 139 days. The example plot of nine features from accelerometer 1 is presented in Figure 4. It can be seen from Figure 4 that not all features represent the degradation condition of slew bearing. Kurtosis, variance and histogram lower features are more sensitive to the bearing condition among the other 9 features. Focusing on the RMS, variance, kurtosis and histogram upper and histogram lower feature that shows a sudden peak on day 90. This is due to the coal dust has inserted to the bearing and makes the roller and raceway has an incipient defect.

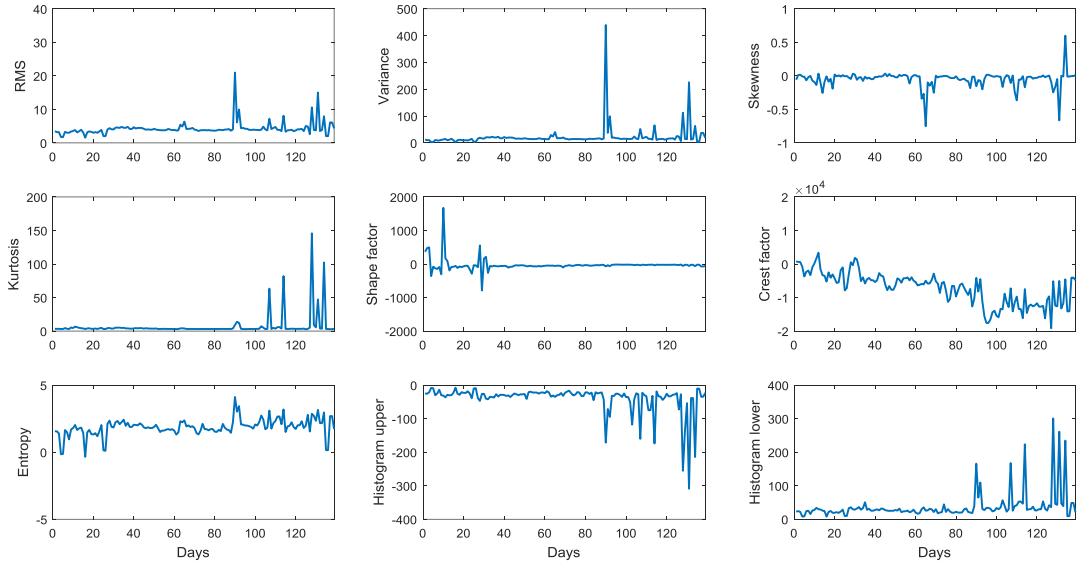

**Figure 4.** Time-domain feature extraction results (139 days).

## 4. Parsimonious network based on fuzzy inference system (PANFIS)

This section presents the working principle of PANFIS [11]. Overview of PANFIS learning policy is outlined in algorithm 1. PANFIS is a type of Evolving Intelligent System (EIS) where it features a fully open structure. It initiates its learning process from scratch and its fuzzy rules can be automatically generated and pruned on demand. PANFIS characterizes a generalized ellipsoidal cluster in high dimension which arbitrarily rotates to any direction. Furthermore, PANFIS is equipped with a fuzzy set extraction strategy which allows crafting fuzzy set representation to obtain transparent traditional IF-Then rules of Takagi Sugeno Kang (TSK) fuzzy system. Fuzzy set merging scenario is incorporated to merge similar fuzzy sets. The rule consequent is constructed with a first-order linear function adjusted using extended recursive least square (ERLS) method [8].

4.1 Network architecture of PANFIS

The unique property of PANFIS's network architecture is observed in the presence of multivariate Gaussian function. The multivariate Gaussian function possesses a non-diagonal covariance matrix which triggers a non-axis-parallel ellipsoidal cluster [11]. Compared to the traditional Gaussian function with the standard product *t-norm* operator, such cluster offers better coverage to irregular data distribution which does not span in the main axes. As a result, it suppresses the fuzzy rule requirement to a lower level than that the standard Gaussian rule. The multivariate Gaussian rule is, however, less transparent than the standard Gaussian rule since the atomic clause of the IF-Then rule (AND-part) is omitted and thus does not have a fuzzy set representation. This calls for a transformation strategy which makes possible to form accurate representation of a fuzzy set. The rule consequent of PANFIS is constructed under the first order TSK fuzzy rule which features a linear function in the output space. The multivariate Gaussian function producing the rule firing strength is expressed as follows:

$$R_i = \exp(-(X - C_i)\Sigma_i^{-1}(X - C_i)^T) \tag{1}$$

where $X_n \in \Re^{1 \times n}$ is the input feature of interest and $C_i \in \Re^{1 \times u}$ is the center of multivariate Gaussian function while $\Sigma_i^{-1} \in \Re^{u \times u}$ denotes the inverse covariance matrix which steers the orientation and size of non-axis parallel ellipsoidal cluster. $u$ is the input dimension. $R_i$ stands for the firing strength of the fuzzy rule which also indicates the degree of confidence of the fuzzy rule. To guarantee the partition of unity [11], the normalization of rule firing strength is performed as follows:

$$\varphi_i = \frac{R_i}{\sum_{i=1}^{C} R_i} \tag{2}$$

where $C$ is the number of fuzzy rule. The output of PANFIS is resulted from the weighted average of the rule firing strength and the rule consequent as follows:

$$y = \sum_{i=1}^{C} \Omega_i \varphi_i = \frac{\sum_{i=1}^{C} \Omega_i R_i}{\sum_{i=1}^{C} R_i} \tag{3}$$

where $\Omega_i = x_e W_i$ is the rule consequent of PANFIS, while $x_e \in \Re^{1 \times (n+1)}$ and $W_i \in \Re^{(u+1) \times 1}$ are the extended input vector and the output weight vector of the *i-th* rule. The first-order TSK rule consequent is selected here because it offers a higher degree of freedom than the zero-order TSK rule consequent.

4.2 Rule growing mechanism of PANFIS

PANFIS makes use of the concept of datum significance (DS) which estimates the statistical contribution of data streams. A data point offering high statistical contribution leads to a new rule. This concept is inspired by the neuron growing mechanism of GGAP-RBF [9] and SAFIS [7]. The key difference, however, can be found in the fact that PANFIS extends the statistical contribution concept to the framework of multivariate Gaussian function. Suppose that a hypothetical rule (*C*+1) is created using the newest data point, the DS method can be formulated as follows:

$$D_{C+1} = |e_n| \frac{(V_{C+1})^n}{\sum_{i=1}^{C+1}(V_i)^n} \tag{4}$$

where $e_n = |T_n - y_n|$ denotes the system error. $T_n, y_n$ stand for the target value and the predictive output at the *n-th* or current training observation. $V_i$ denotes the volume of *i-th* rule which can be simply calculated as the determinant of the non-diagonal covariance matrix $\det(\Sigma_i)$. It is seen from (4) that a datum is considered to bring significant contribution to the training process provided that it causes a high system error. On the other hand the volume of a rule portrays its zone of influence in the input space and a cluster with small influence zone potentially contributes little during its lifespan. Note that (4) contains the implicit distance information due to the initialization of covariance matrix. This facet addresses the novelty of the hypothetical rule because the hypothetical rule occupying remote region possesses a high coverage span. The hypothetical rule is confirmed as a new rule if it induces high statistical contribution $D_{C+1} \geq g_1$ where $g_1$ denotes a conflict threshold. The higher the values of the conflict threshold the lower the number of fuzzy rules are added, while the lower the values of the conflict threshold the higher the number of fuzzy rules are generated during the training process.

If a new fuzzy rule is added, the center of the multivariate Gaussian function is set as the data sample of interest while the diagonal element of inverse covariance matrix is set according to the ε-completeness principle as follows:

$$C_{C+1} = X_N \tag{5}$$

$$diag(\Sigma_{C+1}^{-1}) = \frac{\max(|C_i - C_{i-1}|, |C_i - C_{i+1}|)}{\sqrt{\ln(\frac{1}{\varepsilon})}} \tag{6}$$

where $\varepsilon$ is usually set at 0.6. The $\varepsilon$-completeness principle borrows the seminal work of DFNN and GDFNN in [72] and [73]. It is said that there does not exist any data point with membership degree less than $\varepsilon$ if this setting is implemented and has been mathematically confirmed. It is worth-noting that the covariance matrix of the multivariate Gaussian function is vital to the success of PANFIS. Too large values lead to averaging while too small values leads to be overfitting.

Another situation may occur during the training process where a data sample induces minor conflict. That is, the rule growing condition is not satisfied $D_{C+1} < g_1$. This triggers the so-called rule premise adaptation phase since these samples remain important to refine the network structure. The original PANFIS utilizes the ESOM to update the winning rule in this situation. The underlying bottleneck of the ESOM for PANFIS rule premise adaptation is seen in the reinversion requirement which sometime causes instability if the covariance matrix is not in the full-rank condition. To correct this shortcoming, the direct update scheme of GENEFIS [6], and GEN-SMART-EFS [5] is adopted. The winning rule is fine-tuned if the rule growing condition is violated as follows:

$$C_{win} = \frac{N_{win}}{N_{win}+1} C_{win} + \frac{(X - C_{win})}{N_{win}+1} \tag{7}$$

$$\Sigma_{win}^{-1} = \frac{\Sigma_{win}^{-1}}{1-\alpha} + \frac{\alpha}{1-\alpha} \frac{(\Sigma_{win}^{-1}(X-C_{win}))(\Sigma_{win}^{-1}(X-C_{win}))^T}{1+\alpha(X-C_{win})\Sigma_{win}^{-1}(X-C_{win})^T} \tag{8}$$

$$N_{win} = N_{win} + 1 \tag{9}$$

where $\alpha = 1/(N_{win}+1)$ and $N_{win}$ is the support of the winning cluster. The update of winning rule also increases the population of the *i-th* cluster. The support of fuzzy rules is involved in the rule premise adaptation scheme in (7), (8) to allow stable adaptation because a cluster will converge when it is occupied by a high number of supports. This, however, calls for the forgetting mechanism in the presence of concept drift to improve sensitivity of a highly populated cluster in accepting new training stimuli. The direct update formula of (8) is obtained from the rank-1 modification principle [5].

4.3 Rule pruning scenario of PANFIS

PANFIS is equipped by a rule base simplification strategy, namely Extended Rule Significance (ERS) method which discards inconsequential rules. That is, inconsequential rules which play little during their lifespan can be detected and in turn pruned. In realm of EIS, the rule pruning strategy is vital to alleviate the risk of overfitting and to improve the interpretability of rule semantics. The rule pruning scenario adopts the same principle of the DS method which approximates the statistical contribution of fuzzy rules. The ERS method is formalized as follows:

$$ERS_i = |\delta_i| \frac{(V_i)^n}{\sum_{i=1}^{C}(V_i)^n} \tag{10}$$

where $\delta_i = \sum_{j=1}^{n+1} W_{i,j}$ stands for the total output weight of the *i-th* rule. The salient feature of the ERS method is in the notion of statistical contribution informs approximation of statistical contribution during the training process under the assumption of uniform distribution. In addition, (10) also measures the contribution of output weights. The fuzzy rule contribution is deemed poor provided it has a low output weight because it results in a small output which can be negligible to the overall predictive output of PANFIS. A fuzzy rule is pruned provided the following condition is met:

$$ERS_i \leq g_2 \tag{11}$$

where $g_2$ is the rule pruning threshold. The higher the value of the rule pruning threshold the higher the number of fuzzy rules are pruned during the training process and vice versa. Because the ERS method and the DS method share similar working principle, $g_1, g_2$ are often selected close to each other.

PANFIS is equipped by the fuzzy set merging strategy which aims to coalesce highly overlapping fuzzy sets. Although two fuzzy rules are well-separated in the high-dimensional space, the overlap in fuzzy sets is usually resulted from the projection to one-dimensional axis. This situation often results in inconsistency of rule semantics because fuzzy rules with similar fuzzy sets generate different rule conclusions. PANFIS utilizes the kernel-based metric principle for fuzzy set merging scenario which compares the center and width of Gaussian fuzzy sets in one joint formula [4]. It is expressed:

$$S_{\text{ker}}(A,B) = e^{-|C_A - C_B| - |\sigma_A - \sigma_B|} \tag{12}$$

This formula holds the following interesting properties.

$$\begin{aligned} S_{\text{ker}}(A,B) = 1 &\Leftrightarrow |C_A - C_B| + |\sigma_A - \sigma_B| \Leftrightarrow C_A = C_B \wedge \sigma_A = \sigma_B \\ S_{\text{ker}}(A,B) < \varepsilon &\Leftrightarrow |C_A - C_B| > \delta \vee |\sigma_A - \sigma_B| > \delta \end{aligned} \tag{13}$$

Two fuzzy sets are merged if the kernel-based metric returns $S_{\text{ker}}(A,B) > 0.8$ as set in [4]. This hyper-parameter controls a tradeoff the intensity of fuzzy set merging scenario where the higher the values the more frequent the fuzzy set merging scenario is executed and vice versa. The fuzzy set merging process itself is carried out as follows:

$$c_{new} = (\max(U) + \min(U))/2$$
$$\sigma_{new} = (\max(U) - \min(U))/2 \tag{14}$$

where $U = \{c_A \pm \sigma_A, c_B \pm \sigma_B\}$. This formula is meant to perform the exact merging in accordance to their α-cuts. It is expected that merged fuzzy sets to deliver the membership degree around 0.6 which ensures the ε-completeness at 0.6. It reflects the inflection points of Gaussian fuzzy set $c \pm \sigma$.

4.4 Fuzzy set extraction and merging scenario

Although PANFIS operates in the high-dimensional space, it is fitted with the fuzzy set extraction scenario which derives the fuzzy set representation of non-axis parallel ellipsoidal cluster. This mechanism allows the classical-interpretable IF-Then rule. The centre of the multivariate Gaussian function can be simply applied to the fuzzy set level, while the radii of fuzzy set should be carefully determined because the ellipsoidal cluster rotates to any direction. There exist two strategies to elicit the fuzzy set radii of multivariate Gaussian function: 1) the first method elicits the eigenvalue and eigenvector of non-diagonal covariance matrix. The disadvantage of the first approach is in its prohibitive computational cost because of the eigenvalue and eigenvector computation, although it offers accurate approximation; 2) the second method enumerates a distance from the center to the cutting point of the ellipsoids in the main axes. This method is computationally light although it is rather inaccurate if the ellipsoidal cluster rotates around 45°. In such circumstance, this method generates too tiny spread. We only focus on the second method here. It is expressed as follows:

$$\sigma_i = \frac{r}{\sqrt{\Sigma_{ii}}} \tag{15}$$

where $r, \Sigma_{ii}$ are respectively the Mahalanobis distance and the diagonal elements of the covariance matrix. This mechanism is executed once completing the training process to show the fuzzy rules to operators.

4.5 Adaptation of rule consequent

PANFIS utilizes the extended recursive least square (ERLS) method to update the rule consequent of the fuzzy rule. This approach differs from the original RLS method because of the insertion of a constant $\alpha$ to improve asymptotic convergence of the weight vector. This approach is inspired by the work in [8] where the system error convergence and the weight vector convergence have been mathematically proven with the help of the Lyaponov stability criterion. The constant $\alpha$ behaves like the binary function where it is activated when the approximation error $\acute{e}$ is greater than the system error $e$. It is worth mentioning that the approximation error refers to the system error before the adaptation process where PANFIS aims to make one-step ahead prediction while the system error corresponds to after the tuning process $e_n = |T_n - y_n|$. In other words, the weight vector remains unchanged if the approximation error happens to be lower than the system error.

$$\alpha = \begin{cases} 1, |\acute{e}| \geq |e| \\ 0, otherwise \end{cases} \quad (16)$$

$$L = Q_i x_e (\Psi_i^{-1} + x_e^T Q_i x_e)^{-1} \quad (17)$$

$$Q_i = (I - \alpha L x_e^T) Q_i \quad (18)$$

$$W_i = W_i + \alpha L (t - x_e^T W_i) \quad (19)$$

where $Q_i, L$ are the covariance matrix of the *i-th* rule and the Kalman gain respectively. For the global learning scenario, the covariance matrix $Q$ is global where it embraces the covariance matrix of all fuzzy rules. Since PANFIS is evolving in nature – fuzzy rules can be dynamically added -, rule consequent is set as $W_{C+1} = W_{winner}$ if a new rule is introduced. This setting reflects the fact where the winning rule represents the most adjacent region to the new rule. On the other hand, the covariance matrix of a new rule is set as $Q_i = \alpha I$ in the case of local learning, whereas it is assigned in the same way for the global learning case but the dimension of the covariance matrix have to be expanded. Note that $\alpha$ must be set as a large positive constant to assure convergence of predictive model. ERLS method can be also seen as a variation of FWRLS method in [3] where the binary function is used to enhance the convergence of adaptation process.

## 5. Time-series feature prediction

This section presents our numerical study on the slew bearing prognosis method using the 139 data samples which corresponds to 139 daily records of vibration signals. Nine input features, namely RMS, variance, skewness, kurtosis, shape factor, crest factor, entropy, histogram upper and histogram lower, are extracted. Our simulation was carried out under two modes: direct and time-series.

5.1 Direct mode prediction

The direct mode prediction aims to study the correlation among input features, where 8 features are used as the input attributes to predict the trend of input attributes. For instance, the kurtosis is made as the target variable while the other 8 variables serve as the input attributes to guide the predictive analytics. This simulation aims to illustrate descriptive power of input attributes to indicate the degradation of the slew bearing. Another goal is to model the characteristic of an input attribute based on its nonlinear relationship with other 8 input features. The 139 data points are partitioned into two groups: training and testing where 108 samples are reserved for the training samples, while the rests are fed to test the model. In other words, prediction or testing is done for one-month period. This scenario was carried out for all 9 input attributes. PANFIS's performance is examined based on its generalization power for the testing samples. Figure 5 pictorially exhibits predictive trend of PANFIS in modelling the kurtosis feature under the direct mode. It is shown that the kurtosis feature characterizes nonlinear, uncertain and non-stationary nature but all of which are learnt well by PANFIS where accurate prediction can be resulted from. It is worth noting that under the direct mode one can observe generalization power of PANFIS because the model is fixed after 108 days.

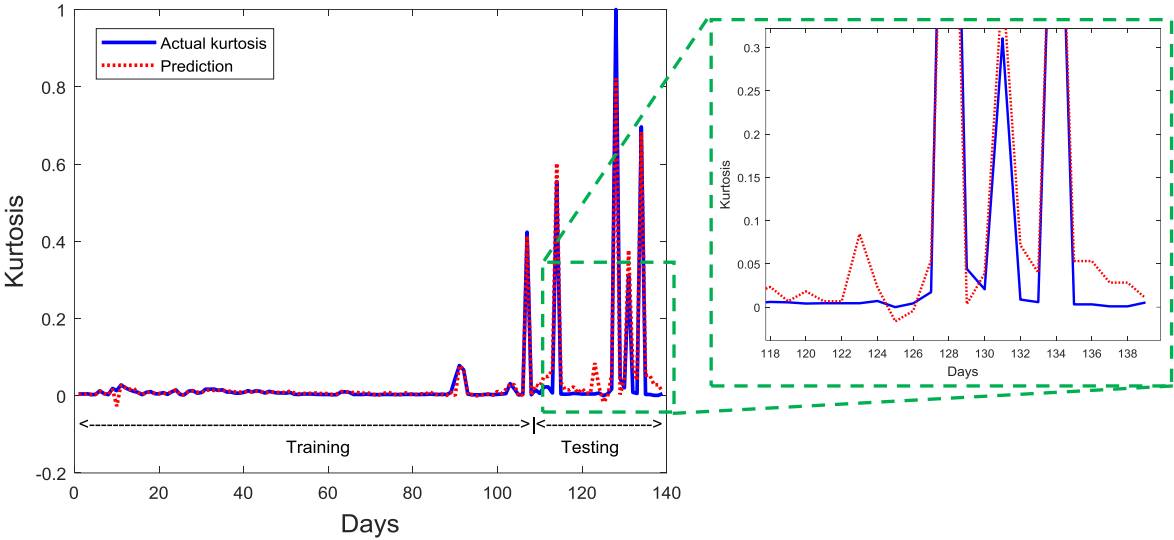

**Figure 5.** PANFIS prediction of kurtosis feature

The evolving and adaptive characteristic of PANFIS is illustrated in Figure 6 where it shows the fuzzy rule evolution of PANFIS in the kurtosis feature problem. PANFIS starts its learning process from no rule at all until the fuzzy rules are automatically created and pruned on the fly in accordance to the novelty of data streams. PANFIS responds timely changing characteristics of the system where it introduces a new rule at *t=88* when there exists "spike" in the kurtosis feature.

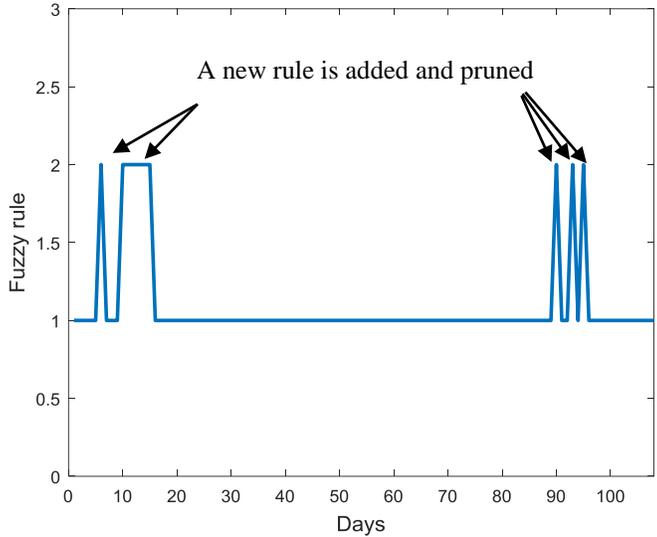

**Figure 6.** Fuzzy rule evolution of kurtosis feature

5.2 Time-series mode prediction

The time-series mode aims to deliver one step ahead prediction based on the previous two consecutive data samples $y_n = f(y_{n-1}, y_{n-2})$. The progression of the slew bearing wear is traced by the predictive model. All samples are fed to the training process and the predictive quality is evaluated by the training error which pinpoints to what extent the predictive model learns given training representation.

PANFIS is compared against three prominent algorithms, eTS [3], simp_eTS [9] and ANFIS [2]. eTS and simp_eTS are counterparts of PANFIS which features both structural and parameter learning scenarios in the online manner. This comparison is needed to confirm the learning performance of PANFIS with respect to similar algorithms. ANFIS is a pioneer of fuzzy neural network (FNN) which occurs to be more traditional than PANFIS, eTS and simp_eTS. It adopts an offline learning scenario where the training process is repeated over multiple epochs. Ten epochs are applied in our study. This comparison aims to demonstrate that although PANFIS works fully in the one-pass learning scenario, it produces comparable predictive quality. Table 3 and 4 present the numerical results of consolidated algorithms in both time-series and direct modes.

It is evident from Table 1 that PANFIS outperforms the other three algorithms in terms of predictive quality. Although ANFIS is reported to have lower RMSE in few cases, it experienced 10 epochs of the training process. Also, we found that ANFIS suffers from the curse of dimensionality notably when the grid partitioning method is used. Hence, we fix to 2 numbers of rule in our simulations. The advantage of PANFIS is more obvious in the direct mode than that in the time-series mode as displayed in Table 2. It outperforms other three algorithms in terms of RMSE, rule and fuzzy set in almost all study cases.

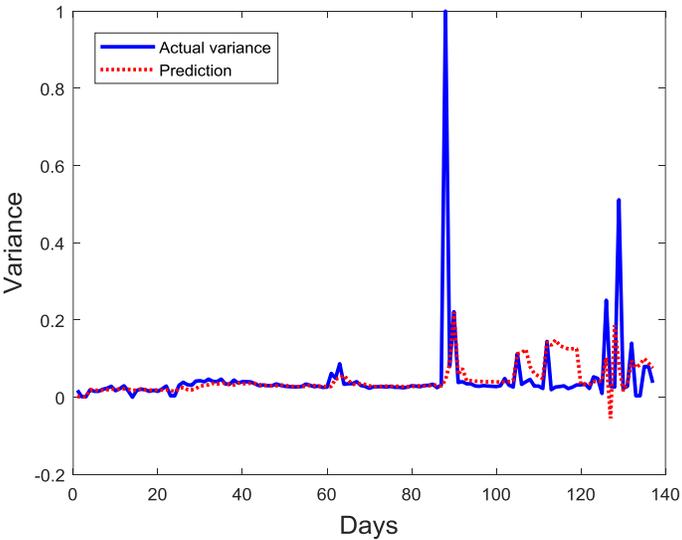

**Figure 7.** PANFIS prediction of variance feature

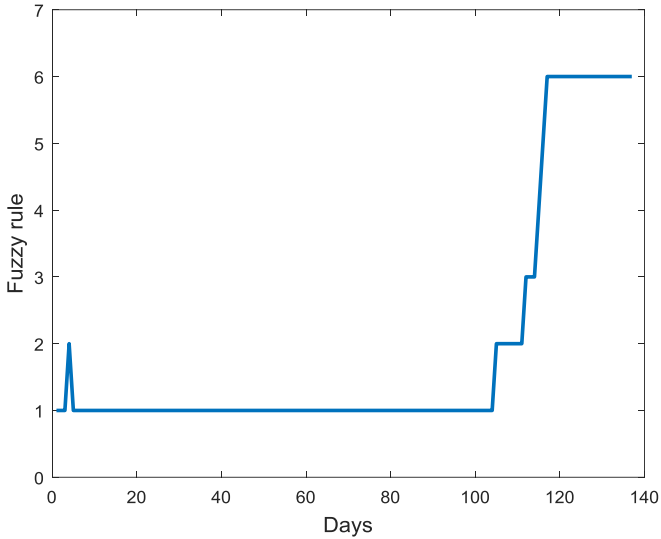

**Figure 8.** Fuzzy rule evolution of variance feature

Figure 7 visualizes PANFIS's prediction of variance feature under the time-series mode. It is seen that PANFIS perfectly models the variance attribute using only its previous two time steps. As with Figure 6 and Figure 8 demonstrates the adaptive trait of PANFIS. It keeps changing on demand which follows nonlinear and time-varying properties of the given problem. PANFIS is also capable of crafting a classical-transparent fuzzy rule from high-dimensional ellipsoidal rule through fuzzy rule transformation (13). An example of PANFIS's rule is exemplified in variance prediction:

$$R_1 : \textbf{IF} \quad Variance = \begin{bmatrix} Variance_{n-1} \\ Variance_{n-2} \end{bmatrix} is \quad close \quad to \quad C_1 = \begin{bmatrix} 0.290 \\ 0.292 \end{bmatrix}, \Sigma_1^{-1} = \begin{bmatrix} 7.4, 0.19 \\ 0.19, 7.4 \end{bmatrix} \textbf{Then}$$
$$y = 0.03 + 0.17 \, variance_{n-1} + 0.04 \, variance_{n-2} \quad (20)$$

This fuzzy rule is rather vague and is unable to be associated directly with linguistic labels because of the absence of atomic clauses. This issue can be addressed using the fuzzy set transformation strategy (13) and this leads to the traditional expression of fuzzy rule as follows:

$$R_1 : \textbf{IF } Variance_{n-1} \textbf{ is close to } c_{11} = 0.29, \sigma_{11} = 0.11 \, Skewness_{n-2} \textbf{ is close to } c_{12} = 0.292, \sigma_{12} = 0.11 \textbf{ Then}$$
$$y = 0.03 + 0.17 \, variance_{n-1} + 0.04 \, variance_{n-2} \quad (21)$$

This rule is more readable than (18) because each fuzzy set corresponds to specific linguistic label. This fuzzy rule is generated under the time-series mode.

Table 3. Numerical results using time-series scenario

| Input Feature | | PANFIS | eTS | Simp_eTS | ANFIS |
|---|---|---|---|---|---|
| RMS | RMSE | **0.1** | 0.11 | 0.13 | **0.1** |
| | Rule | **2** | 5 | 2 | **2** |
| | Fuzzy Set | **2-2** | 5-5 | 2-2 | **2-2** |
| | Time | 0.73 | **0.54** | 1.54 | N/A |
| Variance | RMSE | **0.09** | 0.11 | 0.14 | **0.09** |
| | Rule | 6 | 4 | **2** | **2** |
| | Fuzzy Set | 6-5 | 4-4 | **2-2** | **2-2** |
| | Time | 1.1 | **0.17** | 1.36 | N/A |
| Skewness | RMSE | **0.08** | 0.09 | 0.09 | 0.12 |
| | Rule | **2** | 5 | 4 | **2** |
| | Fuzzy Set | **2-2** | 5-5 | 4-4 | **2-2** |
| | Time | 0.78 | **0.12** | 1.42 | N/A |
| Kurtosis | RMSE | 0.13 | 0.13 | 0.18 | **0.11** |
| | Rule | 16 | 6 | 3 | **2** |
| | Fuzzy Set | 16-16 | 6-6 | 3-3 | 2-2 |
|  | Time | 0.49 | **0.09** | 1.72 | N/A |
| Shape factor | RMSE | 0.11 | 0.09 | 0.09 | **0.06** |
|  | Rule | 4 | 5 | **2** | **2** |
|  | Fuzzy Set | 4-3 | 5-5 | **2-2** | **2-2** |
|  | Time | 0.55 | **0.17** | 1.4 | N/A |
| Crest factor | RMSE | **0.09** | 0.14 | 0.14 | 0.12 |
|  | Rule | 3 | 6 | 7 | **2** |
|  | Fuzzy Set | 3-3 | 6-6 | 7-7 | **2-2** |
|  | Time | 0.49 | **0.15** | 2.1 | N/A |
| Entropy | RMSE | **0.11** | 0.14 | 0.15 | **0.11** |
|  | Rule | 6 | 6 | 5 | **2** |
|  | Fuzzy Set | 6-6 | 6-6 | 5-5 | **2-2** |
|  | Time | 0.5135 | **0.12** | 1.81 | N/A |
| Histogram upper | RMSE | 0.13 | 0.15 | 0.16 | **0.12** |
|  | Rule | 7 | 8 | 6 | **2** |
|  | Fuzzy Set | 7-7 | 8-8 | 6-6 | **2-2** |
|  | Time | 0.69 | **0.19** | 1.7 | N/A |
| Histogram lower | RMSE | **0.14** | 0.16 | 0.2 | **0.14** |
|  | Rule | **2** | 8 | 5 | **2** |
|  | Fuzzy Set | **2-2** | 8-8 | 5-5 | **2-2** |
|  | Time | 0.52 | **0.15** | **1.5** | N/A |

Table 4. Numerical results using direct partition scenario

| Input Feature |  | PANFIS | eTS | Simp_eTS | ANFIS |
|---|---|---|---|---|---|
| RMS | RMSE | **0.03** | 0.04 | 0.25 | 0.32 |
|  | Rule | **1** | 5 | 2 | 2 |
|  | Fuzzy Set | **8** | 40 | 16 | 16 |
|  | Time | 0.41 | **0.12** | 0.15 | N/A |
| Variance | RMSE | **0.04** | **0.04** | 0.21 | 0.11 |
|  | Rule | **1** | 5 | 2 | 2 |
|  | Fuzzy Set | **8** | 40 | 16 | 16 |
|  | Time | 0.43 | **0.13** | 0.15 | N/A |
| Skewness | RMSE | **0.11** | 0.21 | 0.25 | 0.33 |
|  | Rule | 8 | 5 | **2** | **2** |
|  | Fuzzy Set | 64 | 40 | **16** | **16** |
|  | Time | 0.61 | **0.09** | 0.15 | N/A |
| Kurtosis | RMSE | **0.05** | 0.27 | 0.43 | 0.09 |
|  | Rule | **1** | 5 | 2 | 2 |
|  | Fuzzy Set | **8** | 40 | 16 | 16 |
|  | Time | 0.49 | **0.13** | 1.1 | N/A |
| Shape factor | RMSE | **0.006** | 0.05 | 0.11 | 0.33 |
|  | Rule | **1** | 5 | 2 | 2 |

|  | Fuzzy Set | **8** | 40 | 16 | 16 |
|  | Time | 0.52 | **0.19** | 1.36 | N/A |
| Crest factor | RMSE | **0.14** | 0.23 | 0.79 | 8.69 |
|  | Rule | 5 | 5 | **2** | **2** |
|  | Fuzzy Set | 40 | 40 | **16** | **16** |
|  | Time | 0.35 | **0.1** | 1.29 | N/A |
| Entropy | RMSE | **0.14** | 0.22 | 0.79 | 0.15 |
|  | Rule | 5 | 5 | **2** | **2** |
|  | Fuzzy Set | 39 | 40 | **16** | **16** |
|  | Time | 0.64 | **0.13** | 1.4 | N/A |
| Histogram upper | RMSE | **0.08** | 0.3 | 0.28 | 0.10 |
|  | Rule | **1** | 5 | 2 | 2 |
|  | Fuzzy Set | 8 | 40 | **16** | **16** |
|  | Time | 0.57 | **0.15** | 1.01 | N/A |
| Histogram lower | RMSE | **0.05** | 0.37 | 0.35 | 0.54 |
|  | Rule | **1** | 5 | 2 | 2 |
|  | Fuzzy Set | **8** | 40 | 16 | 16 |
|  | Time | 0.4 | **0.15** | 1.2 | N/A |

## 6. Conclusions

A number of literature review on prognosis methods have been found in decades and each review paper presented their own terminology of prognosis approaches. This paper aims to consolidate all the prognosis approaches to provide a clearer understanding and guidelines in selecting the particular prognosis approach for specific needs. Following the review, a study on the data-driven prognosis approach has also been presented in this paper. PANFIS based method, which is considered as one of data-driven prognosis approaches is developed and presented in this paper. PANFIS provides a solution for online prognostic requirement where it characterizes a fully open structure and operated in the single-pass learning mode. This trait makes possible to handle non-stationary characteristics of the system in the sample-wise manner. PANFIS is tested in run-to-failure low speed slew bearing vibration data to predict the time domain vibration features. A comparison study of PANFIS to other three prominent algorithms such as eTS, simp_eTS and ANFIS is also presented. It is shown that PANFIS offers a better prediction performance compared to the three methods.

# Acknowledgement

The data used in this paper were collected from Tribology Laboratory, School of Mechanical, Materials, Mechatronic and Biomedical Engineering, University of Wollongong, Australia. The first author thanks the University of Wollongong, Australia for the financial support through International Postgraduate Research Scholarship (IPRS) during the test rig construction and experimental setup. The second author acknowledges the support of Nanyang Technological University start-up grant and MOE Tier-1 grant.